\documentclass[journal,10pt]{IEEEtran}
\usepackage{cite,graphicx,amsmath,amssymb}
\usepackage{subfigure}
\usepackage{citesort}
\usepackage{fancyhdr}
\usepackage{mdwmath}
\usepackage{mdwtab}
\usepackage{balance}
\usepackage{xcolor}
\usepackage{bm}
\usepackage{amsthm}
\usepackage{algorithm}
\usepackage{algorithmic}
\usepackage{multirow}
\usepackage{flafter}
\usepackage{threeparttable}

\begin{document}
\title{A New Look at AI-Driven NOMA-F-RANs: Features Extraction, Cooperative Caching, and Cache-Aided Computing}
\markboth{\textit{A Manuscript Submitted to The IEEE Wireless Communications} }{}

\date{\today}
\author{
Zhong~Yang,~\IEEEmembership{Student Member,~IEEE,}
Yaru~Fu,~\IEEEmembership{Member,~IEEE,}
Yuanwei~Liu,~\IEEEmembership{Senior Member,~IEEE,}
Yue~Chen,~\IEEEmembership{Senior Member,~IEEE,}
Junshan Zhang,~\IEEEmembership{Fellow,~IEEE}

\thanks{Zhong~Yang, Yuanwei~Liu and Yue~Chen are with Queen Mary University of London, London,
UK (email: \{zhong.yang,yuanwei.liu, yue.chen\}@qmul.ac.uk).}
\thanks{Yaru~Fu is with the School of Science and Technology, Hong Kong Metropolitan University, Hong Kong, 999077 China (Email: yfu@hkmu.edu.hk).}
\thanks{Junshan~Zhang is with School of Electrical, Computer and Energy Engineering, Arizona State University, Tempe, AZ (e-mail: Junshan.Zhang@asu.edu).}}
 \maketitle

\begin{abstract}

Non-orthogonal multiple access (NOMA) enabled fog radio access networks (NOMA-F-RANs) have been taken as a promising enabler to release network congestion, reduce delivery latency, and improve fog user equipments' (F-UEs') quality of services (QoS). Nevertheless, the effectiveness of NOMA-F-RANs highly relies on the charted feature information (preference distribution, positions, mobilities, etc.) of F-UEs as well as the effective caching, computing, and resource allocation strategies. In this article, we explore how artificial intelligence (AI) techniques are utilized to solve foregoing tremendous challenges. Specifically, we first elaborate on the NOMA-F-RANs architecture, shedding light on the key modules, namely, cooperative caching and cache-aided mobile edge computing (MEC). Then, the potentially applicable AI-driven techniques in solving the principal issues of NOMA-F-RANs are reviewed. Through case studies, we show the efficacy of AI-enabled methods in terms of F-UEs' latent feature extraction and cooperative caching. Finally, future trends of AI-driven NOMA-F-RANs, including open research issues and challenges, are identified.
\end{abstract}

\vspace{-0.3cm}
\section{Introduction}

The past years have witnessed an explosive growth of intelligent mobile services, including augmented reality, virtual reality, holographic telepresence, industry 4.0, and robotics, benefiting from the rapid upgrades of terminal/wireless communication techniques as well as the emerging artificial intelligent (AI) technologies. These innovative applications are in general latency-sensitive and massive data-driven, which impose great pressure to the conventional cloud radio access networks (C-RANs) due to its centralized processing mode and limited fronthaul capacity. To address these issues, the focus of the research community is shifting towards the system design of fog radio access networks (F-RANs)~\cite{Zhou2019pieee,Mugen2016inetwork}. In F-RANs, two significant changes are made compared to C-RANs, i.e., proactive caching and computing capabilities are enabled at the fog access points (F-APs), which bring a two-fold benefit. On one hand, the frequently requested contents of F-UEs can be pre-fetched and cached at F-APs during off-peak time, resulting in a decreased delivery latency of F-UEs. On the other hand, the computational-intensive and latency-sensitive tasks of F-UEs can be executed at F-APs via task offloading, which significantly decrease the energy consumption of mobile F-UEs. Moreover, the tasks' results can be cached at F-APs, and are downloaded by other F-UEs directly for further enhancing the computing performance. The foregoing aspects, in turn, release the heavy transmission workload of both backhaul and frounthaul links.

Meanwhile, non-orthogonal multiple access (NOMA) is being considered as another promising enabler for next-generation wireless networks due to its high spectrum efficiency and ultra-high connectivity. Different from traditional orthogonal multiple access (OMA) techniques that serving different users in orthogonal resource blocks, NOMA techniques have shown great potential for circumventing the limitation of massive connectivity. It is anticipated that the unprecedented high access speed and low latency requirements in F-RANs could be well solved by the NOMA techniques. The integration of NOMA and F-RANs has given rise to a new research area, namely, ``NOMA-F-RANs''. In NOMA-F-RANs, NOMA techniques are conducted for the content delivering from the base station (BS) to mobile users and task offloading when mobile user request for the computational tasks.

\begin{figure*} [!t]
\setlength{\abovecaptionskip}{-0.2cm}
  \centering
  \includegraphics[width=3.4in]{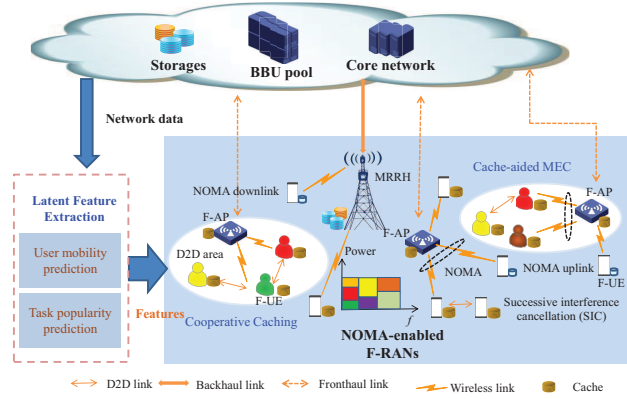}\\
  \caption{Network structure of NOMA-F-RANs}\label{structure}
\end{figure*}

\subsection{Related Work}

Considering the potential advantages of NOMA-F-RANs in next-generation wireless networks, NOMA-F-RANs have received widespread attention for addressing the challenging data rate and latency requirements of F-UEs. From the transmission perspective, in contrast to conventional OMA techniques, applying the NOMA technique in F-RANs is capable of better utilizing the communication capacity for content delivering and computational tasks offloading. From the computation perspective, tasks' offloaded computing and content caching are efficient approaches to reduce energy consumption and computing latency for F-UEs. Previous studies on NOMA-F-RANs focus on problems, such as, task offloading decision, computing resources allocation, user clustering, and power allocation. Conventional optimization approaches, such as branch-and-bound (BnB) algorithm, heuristic algorithm, and matching game algorithm have been applied to solve the aforementioned problems. However, optimization based techniques in general requires a large number of iterations to find a satisfying solution, which is unaffordable for achieving a real-time optimization in fast-fading wireless communication systems. The recent advances in AI offer promising approaches to tackle the aforementioned challenges. For instance, deep learning (DL) algorithms have strong fitting capability, which can be applied to extract complex hidden information from big data collected from the network~\cite{Ramesh2017aaai}. Furthermore, reinforcement learning (RL) algorithms can be utilized for stochastic optimization problems of task offloading and computing resources allocation, which may not computationally feasible for conventional optimization approaches. For brevity, a summary of some recent research contributions of the AI-based algorithms on NOMA-F-RANs is provided in Table~\ref{tableaifran} with the highlights of the adopted AI solutions as well as the objectives.

\subsection{Motivation}

Clearly, for stochastic optimization problems in NOMA-F-RANs, the certainty and availability of prior feature information such as user mobility and content popularity highly affect system's effectiveness. While some AI-enabled solutions have been developed for solving cache decision or resource management problems, the F-UEs' feature information are stipulated to be known as prior~\cite{Guo2020twc,Xiong2020jsac,Chen2018tcom,Sadeghi2019jsac,Xiao2018tvt,chaofan2019jsac,Doan2020tcom}. In real systems, the spatial-temporal information is difficult to track and obtain, especially when the number of F-UEs is large. Moreover, the potential gain of cooperative caching and cache-enabled computing in NOMA-F-RANs are not well revealed by existing work, which calls for new research to provide an overarching perspective towards AI-driven NOMA-F-RANs, as proposed by this article. Different form existing paper on NOMA-FRANs~\cite{Zhang2018wmcom}, in this article, we propose an AI-driven NOMA-F-RANs architecture that sheds light on how novel AI techniques can be exploited for addressing the principle issues and enhancing the performance of key modules, namely, cooperative caching and cache-aided mobile edge computing (MEC) in NOMA-FRANs. Specifically, we first present the system structure of the NOMA-F-RANs in Section II, highlighting the key modules. Then, we explain how AI-based solutions can be applied to address the principle issues in the considered structure. Two case studies are provided in Section IV, to show the performance of our versatile AI solutions. Finally, we conclude this paper and point out future research trends in Section V.

\begin{table*}[t!]
\caption{Artificial Intelligent Solutions for NOMA-F-RANs}
    \centering
	\begin{tabular}{|c||c||c||c|}\hline
    	{\bf References}& {\bf Networks~or~framework}&{\bf Objectives}&{\bf Employed~AI~solutions}\\\hline
        \cite{Guo2020twc} & Blockchain-based MEC & Task offloading &  Double-dueling DQN  \\\hline
        \cite{Xiong2020jsac} & IoT edge computing & Computing allocation &  DQN with multiple replay memories  \\\hline
        \cite{Chen2018tcom} & Content caching networks& Content placement & Echo liquid state machine learning  \\\hline
        \cite{Sadeghi2019jsac} & Content caching networks& Fetch-caching decision & Modified online solver based Q-learning \\\hline
        \cite{Xiao2018tvt} & NOMA downlink networks& Power allocation & Hotbooting Q-learning  \\\hline
        \cite{chaofan2019jsac} & NOMA downlink networks & Channel assignment & Attention-based NN enabled DRL  \\\hline
        \cite{Doan2020tcom} & Cache-aided NOMA downlink & Power allocation & Dual-networks enabled DRL \\\hline
        \cite{Yan2020tcom} & Caching in NOMA-FRANs & User Access Mode Selection & Evolutionary game  \\\hline
	\end{tabular}
	\label{tableaifran}
    \begin{tablenotes}
    \footnotesize
    \item DQN represents deep Q-network; IoT represents Internet of Things; NN is the acronym for neural network. DRL represents deep reinforcement learning; D2D is short for device-to-device; IL represents independent learners.
    \end{tablenotes}
\end{table*}

\section{NOMA-F-RANs System Structure}\label{section:nomaenablefrans}
In this section, we first present the NOMA-F-RANs architecture. Then, we discuss the associated key modules.
\vspace{-0.3cm}
\subsection{Network Structure for NOMA-F-RANs}
Fig.~\ref{structure} illustrates the network structure of NOMA-F-RANs. Therein, NOMA technique is adopted for content delivering and task offloading between F-APs and F-UEs. In the meanwhile, fronthaul links connect the baseband units (BBU) pools and F-APs. Morover, fronthaul links also connect BBU pools and the macro remote radio head (MRRH), which has centralized computing and storage capabilities. In F-RANs, the F-APs works from three aspects, namely remote transmission, radio resources management and task offloaded computing for F-UEs. Furthermore, the rapid development of terminal technologies, such as high-performance workstations, has enabled F-APs and the F-UEs in F-RANs much stronger caching and computing capabilities. Thanks to device-to-device (D2D) techniques, F-UEs can request contents/tasks results from both F-APs and nearby F-UEs directly. As a consequence, duplicated task offloaded computing and content delivering from BBU to F-APs can be saved. Both backhaul and fronthaul burden is alleviated accordingly. As can be seen from the Fig. 1, network data is first obtained from the core network for latent feature extraction. Then, the extracted latent features, such as user mobility and content popularity are utilized for resource allocation in cooperative caching and cache-aided MEC modules. Hereinafter, we discuss the principal modules of NOMA-F-RANs.
\vspace{-0.3cm}

\subsection{Key Modules in NOMA-F-RANs}

\begin{figure}
\setlength{\abovecaptionskip}{-0.2cm}
  \centering
  \includegraphics[width=3in]{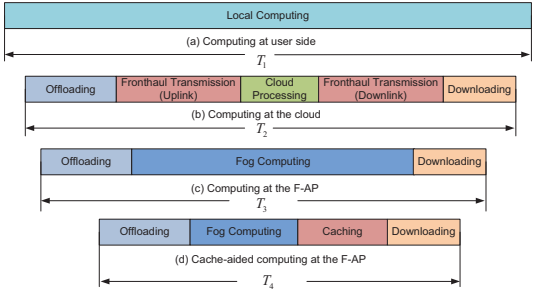}\\
  \caption{Task computing in NOMA-F-RANs}\label{offloading}
\end{figure}

\subsubsection{Latent Feature Extraction}

In NOMA-F-RANs, the joint communication and computing resources allocation requires some prior information of the networks, such as user mobility and task popularity. These latent features need to be extracted according to the history data. However, the task popularity is often assumed following a generalized Zipf law, which is a experience distribution and lacks theoretical guarantee. Advanced AI approaches and social media platforms have attracted more attention in the field of latent feature extraction, by exploiting latent issues in the regions and the interactions among the public, the networks are capable of better predicting the latent features, to reinforce the quality of the wireless network service.

\subsubsection{Cooperative Caching}

Recent research advances on AI-driven content caching focused on non-cooperative caching~\cite{Chen2018tcom,Sadeghi2019jsac}. Although caching at the edge side can be regarded as an efficient manner to alleviate the data traffic and the energy cost of the backhaul links in NOMA-F-RANs, designing the content placement policy for all F-APs independently may insufficiently utilize the caching resources of all the F-APs, because all F-APs attempt to cache the most popular contents. In order to further boost the benefits of caching in NOMA-F-RANs, cooperation among cache storages of different F-APs is of high necessity to be considered. The main ideology of cooperative caching is to place contents collaboratively at all APs, which is capable of increasing the content diversity in the networks. In addition, when considering proactive cache optimizations for F-APs, the similarity among users' content preferences and mobilities should also be taken into account. Since F-UEs from the same social networks or keen on the similar social activities tend to generate the same downlink data streaming. Nevertheless, these prior information (e.g., users' mobilities, users' content preferences, etc) cannot be obtained via conventional optimization based methods.

\subsubsection{Cache-aided MEC}

MEC is a powerful implement for wireless networks to process the computationally-intensive and latency-sensitive tasks for resource limited mobile terminals. Fig.~\ref{offloading} shows the task computing in NOMA-F-RANs. In Fig.~\ref{offloading}(a) and (b), the computational tasks of F-UEs can be computed at the F-UEs side and the cloud, respectively. Moreover, caching task computing results at the MEC server is capable of avoiding duplicate and redundant data transmission, thus streamlining the task offloading decision, reducing the computing latency and saving the task computing energy. The cache-aided MEC for F-UEs in NOMA-F-RANs can be divided into four phases, i.e., task offloading, data processing, task results caching and downloading, as illustrated in Fig.~\ref{offloading}(d), wherein, the computing results are cached in the MEC server. When the same task computing request is arriving, the cached results can be delivered to the F-UEs directly. The computation capability of F-APs is limited. Imaging that the F-APs cannot satisfy the computation requirements, the offloaded tasks will be further sent to BBU pools (cloud) through fronthaul link since in general the cloud server has much greater data processing power than that of the F-APs. After completing all computations, the cloud servers first push back the outcomes to F-APs. Subsequently, F-APs transmit the computational results to their associated F-UEs, as demonstrated in Fig.~\ref{offloading}(c). So far, the hierarchical fog-cloud computing is implemented in NOMA-F-RANs. Fig.~\ref{offloading} outlines the possible computing schemes in NOMA-F-RANs, in which $T_1$, $T_2$, $T_3$, $T_4$, represents the latency requirement of each F-UE. Moreover, since user side has weaker computing capability than the cloud and the F-AP, task computing at the user side should takes more time. If the channel quality is poor, local computing could be the best manner among all. Although the fog-cloud computing saves computational energy at F-UEs, there are additional transmission latency (or fronthaul latency given that cloud computing is adopted) and transmission energy consumption. Therefore, to fully unleash the potential of cache-aided MEC in terms of computation and caching, the optimization of joint offloading decision, caching decision, and the associated radio and computing resources allocation for F-UEs should be well designed. Existing papers applied conventional optimization approaches~\cite{Ying2017lcn} in cache-aided MEC networks to obtain static optimal solution, which is, however, computationally unfeasible when considering long-term caching decisions. Because conventional optimization approaches need to be executed for each time slot. To overcome this drawback, in this article, RL based solutions are proposed for cache-aided MEC networks, which is capable of obtaining a long-term offline policy for task offloading and caching decision. In other words, RL based solution only need to be executed one time, then a long-term policy is obtained for multiple time slots.

\vspace{-0.3cm}

\subsection{Standardization of NOMA-FRANs}
With the swift deployment of 5G wireless networks, there exist some standardization and application efforts on NOMA-FRANs to the envisioned B5G/6G wireless networks from academia, industry, and government agencies. Firstly, NOMA has been included in the 3rd generation partnership project long-term evolution advanced (3GPP-LTE-A) standard, term multiuser superposition transmission (MUST) at 2016. NOMA is also included in the next general digital TV standard (ATSC 3.0), and the 5G new radio (NR) standard. For F-RANs, two service requirements are defined by 3GPP for Release 17 at 2019. Meanwhile, European Telecommunications Standards Institute (ETSI) launched the standardization of F-RANs at 2014. Moreover, although NOMA is exclusive from 5G wireless networks, NOMA-FRANs can be applied for satisfying and enhancing the Quality of Service (QoS) of users, which plays critical effect on novel application envisioned, such as augmented reality (AR), real-time online gaming and high-speed video streaming in B5G/6G wireless networks.

\section{Artificial Intelligent driven NOMA-F-RANs}\label{section:aienablednomafrans}

In this section, we explore how AI can be used to fully obtain the gains of NOMA-F-RANs, which is expanded from two perspectives: latent feature extraction and resource management. The reason for considering latent feature extraction is that massive data collected from social media and network operators stimulate the utilization of novel AI techniques for extracting accurate latent features. Regarding the resource management, RL algorithms have manifest advantages in terms of network optimization and decisions, which can be adopted for resource management in NOMA-F-RANs. Regarding on other perspectives of utilizing AI in NOMA-FRANs, such as user association, and mobility estimation, we will consider them in the future works, employing advanced AI techniques to obtain notable performance enhancement.

\vspace{-0.3cm}
\subsection{Latent Feature Extraction in NOMA-F-RANs}

To efficiently allocate computing and caching resources in NOMA-F-RANs, prior information, e.g., F-UEs' mobility and contents' popularity in the future is required. These latent features cannot be mathematically expressed using conventional approaches, therefore, we advocate to leverage DL to enable NOMA-F-RANs to predict the just-in-time demands of the networks, which is capable of efficiently allocating caching and computing resources, including bandwidth, storage capacity and computing speed.

\begin{figure*}[t!]
    \centering
    \subfigure[RNNs/LSTM networks solution for mobility/task popularity prediction.]{
    \begin{minipage}{6cm}\label{flowLSTM}
    \centering
    \includegraphics[width=6cm]{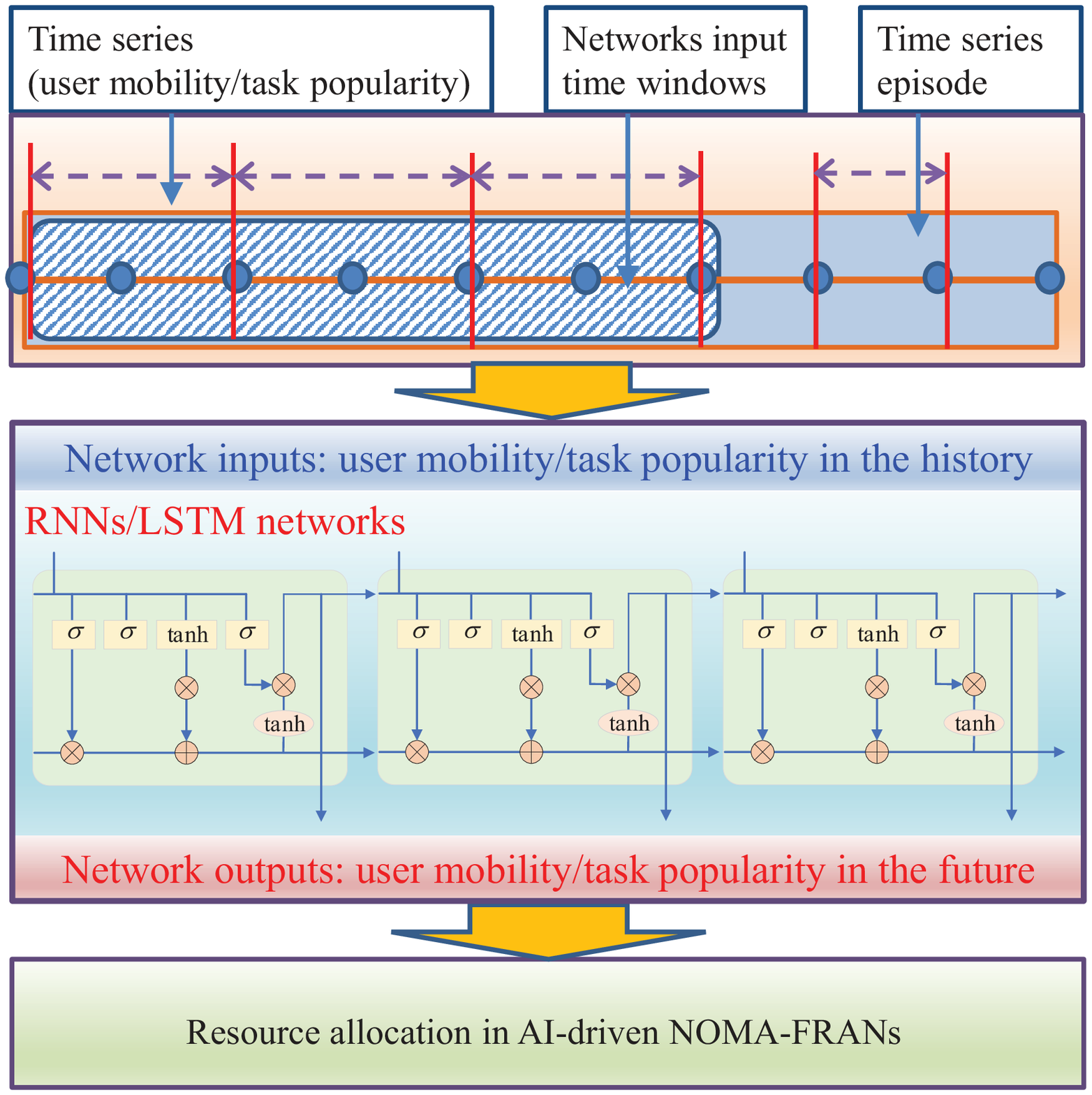}
    \end{minipage}
    }
    \subfigure[LA-based Q-learning for cooperative caching.]{
    \begin{minipage}{6cm}\label{Qlearningflow}
    \centering
    \includegraphics[width=6cm]{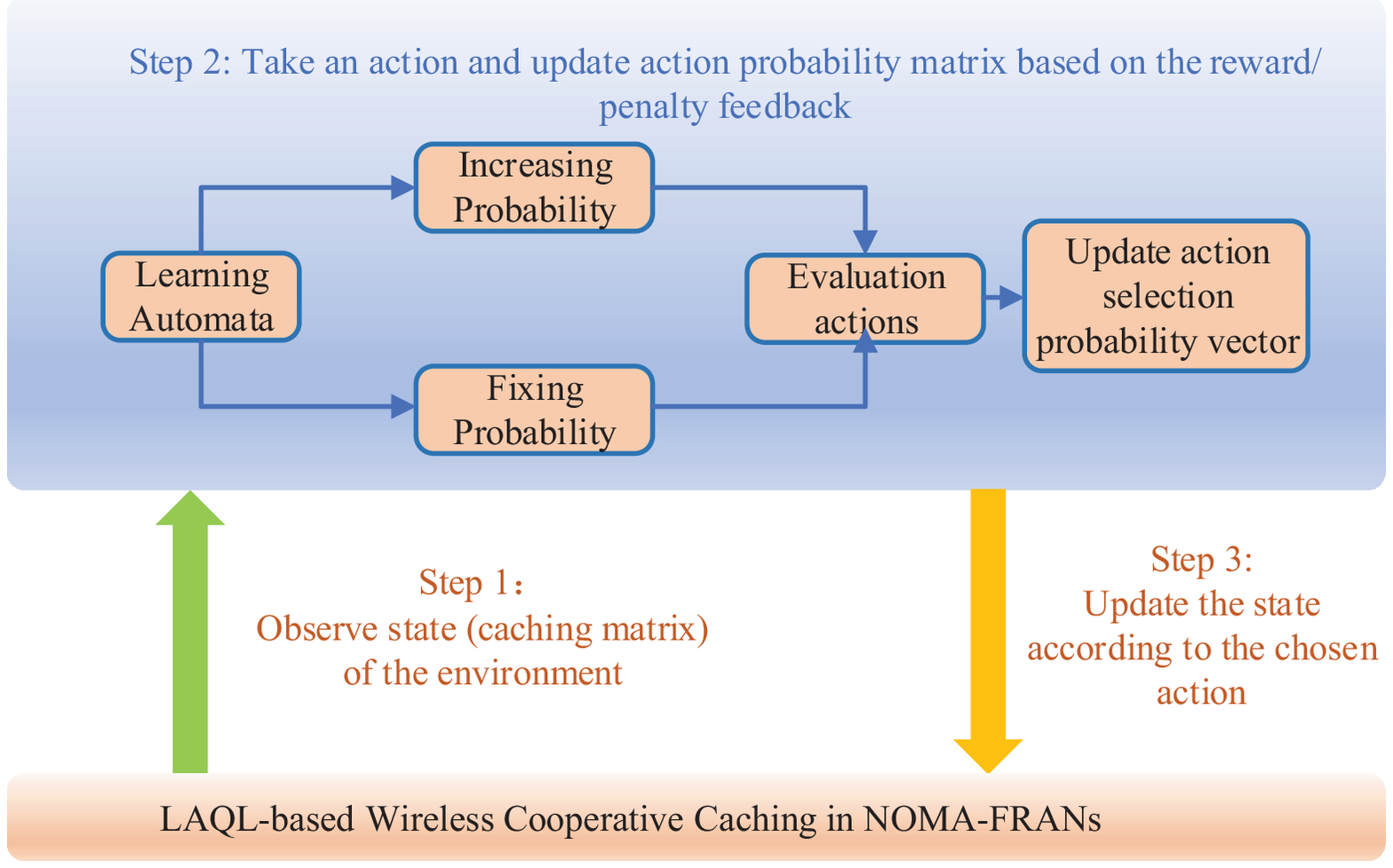}
    \end{minipage}
    }
    \caption{The proposed AI solutions for resource allocation in NONA-FRANs.}
\end{figure*}

Although RNNs have been proposed in previous paper~\cite{Mingzhe2017jwcom}, the investigating on adopting novel RNNs for predicting user mobility and content popularity still require great efforts, especially when a large amount of social media data and network data is generated every day. Because the explosive growth of wireless and social media data is a key driver that boosts AI development and application. With the benefit of massive data collected from social media and network operators, we adopt RNNs and LSTM networks to predict F-UEs' mobility and contents' popularity, respectively, since they are capable of processing time-sequential data due to their internal memory units~\cite{Ramesh2017aaai}. The flow chart of using RNNs and LSTM networks for F-UEs' mobility and contents' popularity prediction is given in Fig.~\ref{flowLSTM}.

\subsubsection{RNNs for User Mobility Prediction}
The mobility of F-UEs plays a critical effect on the performance of cooperative caching and cache-aided MEC in NOMA-F-RANs, because F-UEs with similar social activities tend to produce approximate data flows. Since the mobility of F-UEs can be reshaped as a time series, we use RNNs to predict F-UEs' positions due to its high processing capability for time-sequential data. Universal approximation theorems demonstrate that deep neural networks (DNNs) are capable of fitting complicated functions. Among which, RNNs have been proposed as efficient approaches for solving time series prediction problems~\cite{Ramesh2017aaai}. Henceforth, we invoke RNNs for user position prediction. In particular, the inputs of the RNNs are users' positions in history time slots, while the outputs of the RNNs are the positions of users in the future time slots, which can be used for network resources configuration.
%

\subsubsection{LSTM Networks for Task Popularity Prediction}
To sufficiently provide computing and caching resources for the F-UEs, it is key to obtain the task popularity therein. We obtain the prosperous historical task popularity data from network monitoring with standard manipulation rules or from network operators. Using the obtained dataset, we predict the task popularity by LSTM networks. LSTM networks are widely adopted for time series prediction problems due to the fact that LSTM networks store representations of history input information in form of activations by series connected multiple cells in the networks. The advantages of LSTM networks make them potentially significant for many applications.

In the task popularity prediction problem, a great amount of history data (i.e., the task popularity in the history) is collected as a training dataset. As shown in Fig.~\ref{flowLSTM}, the F-UEs' task popularity is treated as a time series. Therefore, time series prediction algorithms such as RNNs can be utilized for forecast the F-UEs' task popularity in the future. The inputs of the networks are historical data in multiple time slots, and the outputs of the networks are task popularity in the near future. After training, the LSTM networks are adopted to predict the task popularity, which serves as prior information for the resource allocation in the next subsection.

\subsection{AI-driven Joint Resource Allocation in NOMA-F-RANs}

Based on the predicted networks' traffic demands, the prosperity of AI algorithms provide effective and low-cost solutions for NOMA-F-RANs that are capable of fulfilling the stringent requirements adaptive to the F-UEs' mobility and radio environment. We adopt RL in NOMA-F-RANs because the mechanical of RL approaches is obtain a long-term offline policy that maximize the sum weighted reward by balancing exploration and exploitation, which is capable of solving a stochastic optimization problem of joint resources allocation~\cite{chaofan2019jsac,Sadeghi2019jsac}. Since the task offloading decision is a discrete variant in NOMA-FRANs. Therefore, we choose tabular Q-learning and DQN algorithms instead of actor critic, deep deterministic policy gradient (DDPG), or asynchronous advantage actor-critic (A3C) based DRL algorithms, which work for continuous action space and state space. The resource allocation policy is defined by the optimization parameters in NOMA-FRANs, such as content caching decision, task offloading decision, computing resource allocation, and communication resource allocation. As in Fig. 3(b), the state space of RL is defined by the optimization parameters in the considered networks. The action space is the change of the parameters in the state space, which enables state space to cover all the possible combinations of optimization parameters. According to Fig. 3(b), through three learning steps, the RL is capable of given the optimal action for every given state, thus determines the resource allocation policy. Moreover, the structure of LA is simple, thus it is easy to be implemented for action selection of RL. The detail of LA-based action selection is given in Fig. 3(b). Noting that the action selection of LA can only be increased or fixed, which increases the learning speed. The output of LA is action selection probability vector, which is used for updating the Q-table in Q-learning.

\subsubsection{Tabular Q-learning in NOMA-F-RANs}
In NOMA-F-RANs, we focus on the joint computing and caching problem. Moreover, we formulate a long-term stochastic optimization problem that entails a joint optimization of both communication and computing resources for F-UEs. The solution of considered stochastic optimization problem is a long-term offline policy that can be obtained by tabular Q-learning. Tabular Q-learning is one of the model-free RL algorithms that aim to select suitable action to maximize the weighted sum reward in a particular situation by training the Q-table. The reward function of the tabular Q-learning in NOMA-F-RANs is defined by the objective functions in the networks, e.g., energy consumption minimization, summation data rate maximization, computation latency minimization, etc.

\subsubsection{Deep Q network (DQN) in NOMA-F-RANs}

It is widely recognized that conventional value-based RL algorithms suffer from the curse of dimensionality. Especially, in NOMA-F-RANs, the dimensions of state space and action space are settled by the number of network parameters, e.g., number of channels, F-UEs and F-APs. To overcome this drawback, we adopt DQN for the optimization problem in NOMA-F-RANs. In DQN, the optimal policy of the intelligent agent is obtained by updating Q values in NNs. The inputs of the NNs are the current states and the outputs are the probabilities of all the actions in the action space. By utilizing the fitting ability of the NNs, a high-dimension state input and low-dimension action output pattern is implemented to deal with the curse of dimensionality in conventional RL algorithms, especially when the number of network parameters in NOMA-F-RANs are large.

\subsubsection{Learning automata based Q-learning in NOMA-F-RANs}
The aim of learning automata (LA) is to learn the optimal action from the action space for the given network environment. Different from RL algorithms that learn optimal actions for every states, LA is used for an intelligent agent that has only one state. In NOMA-F-RANs, the resource allocation problem entails joint optimization of multiple parameters that are coupled together, which makes it challenge to use RL algorithms for the joint optimization, because it is challenging for intelligent agent to specify reward functions as well as the large variances. To overcome abovementioned challenges, we use LA-enabled Q-learning for joint computing and caching resources allocation problem in NOMA-F-RANs. In the proposed LA based Q-learning algorithm, LA based action selection scheme is proposed for enabling every state to select the optimal action with arbitrary high probability if Q-learning is able to converge to the optimal Q value eventually. The performance of the proposed LA-based Q-learning algorithm compared to conventional Q-learning algorithm in NOMA-F-RANs is shown in next section.

\begin{figure*}[t!]
    \centering
    \subfigure[Task popularity prediction (goal equals to 0.01).]{
    \begin{minipage}{7cm}
    \centering
    \includegraphics[width=7cm]{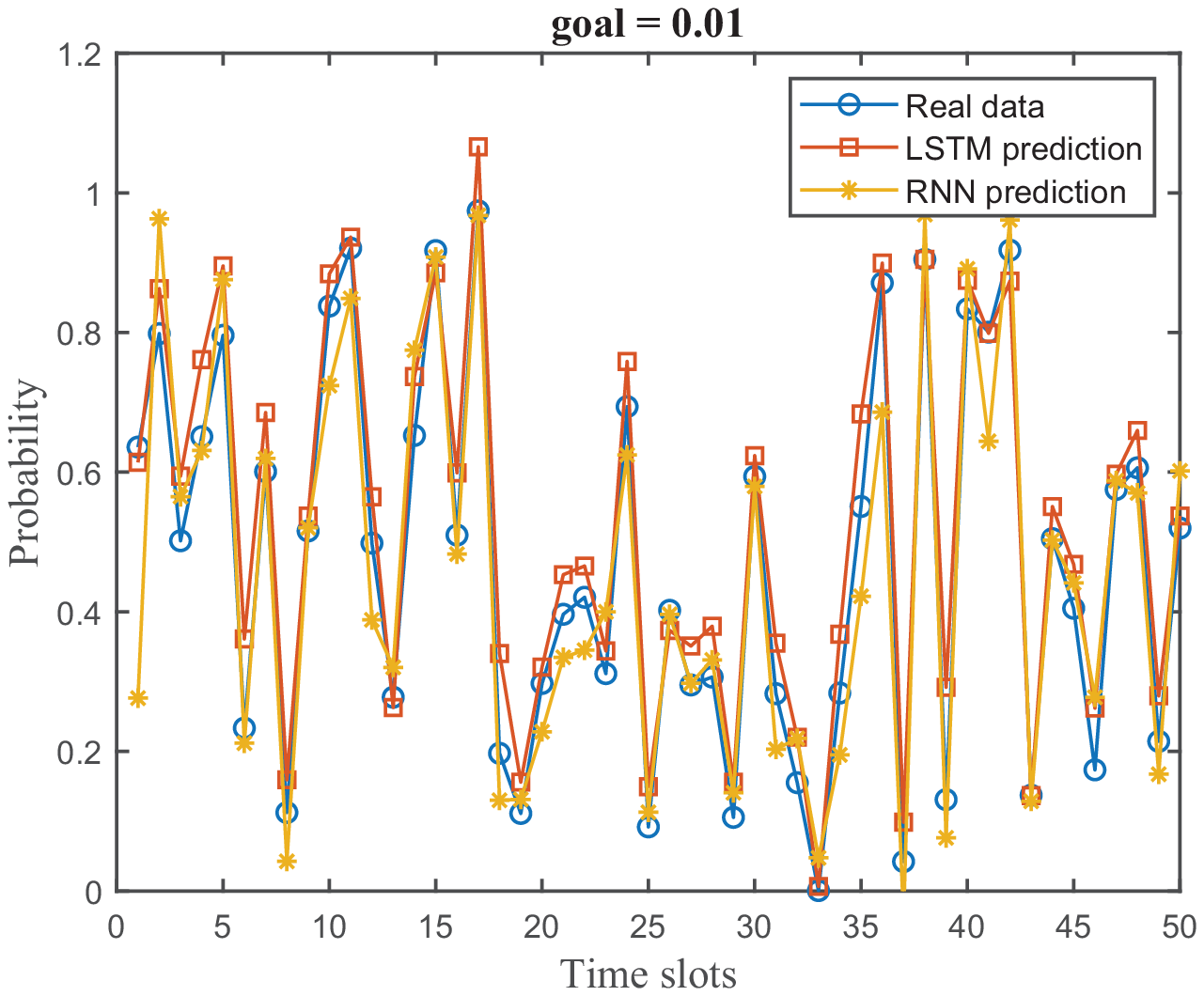}
    \end{minipage}
    }
    \subfigure[Task popularity prediction (goal equals to 0.001).]{
    \begin{minipage}{7cm}
    \centering
    \includegraphics[width=7cm]{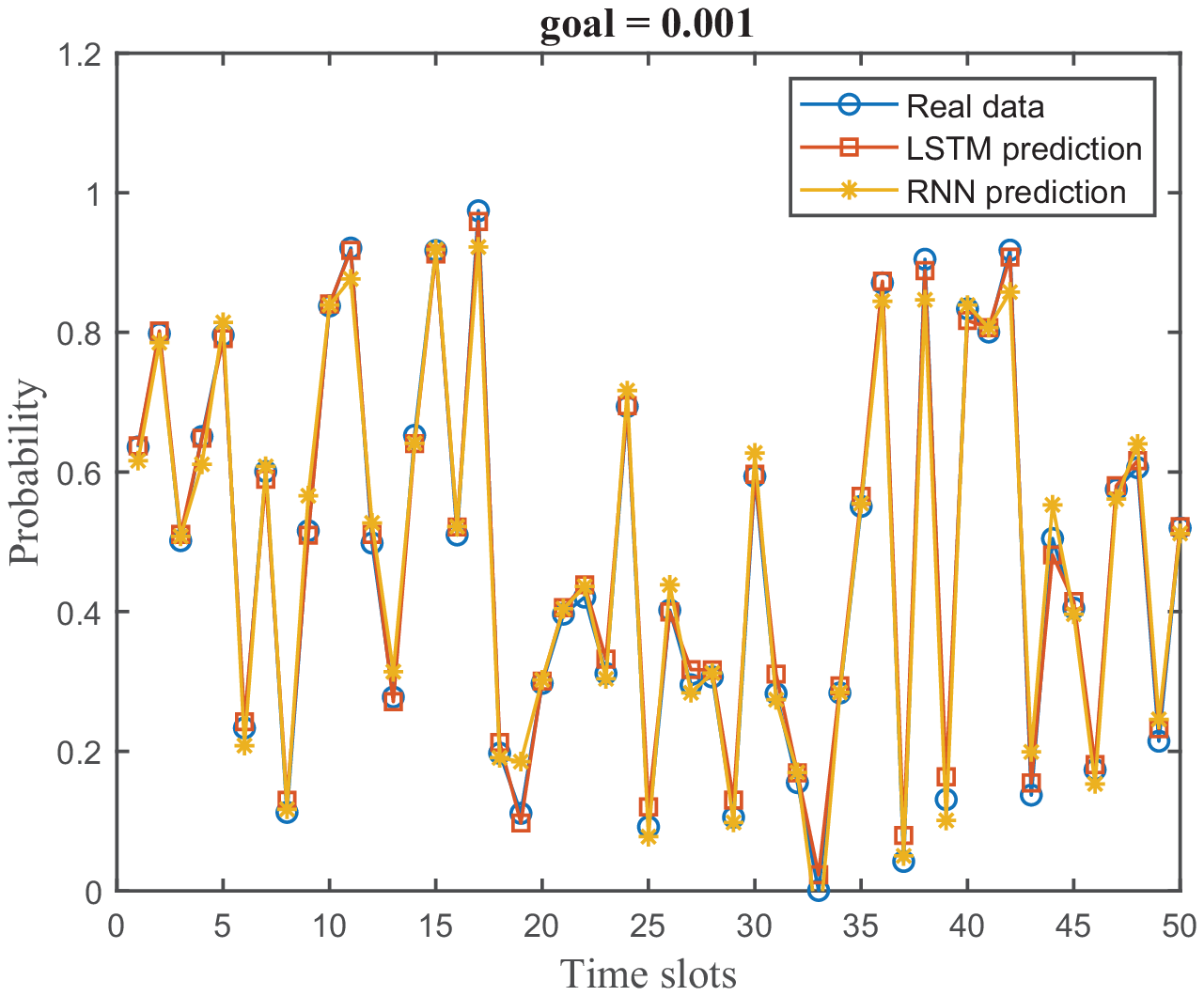}
    \end{minipage}
    }
    \caption{Simulation results of task popularity prediction using LSTMs and RNNs.}\label{performanceofLSTM2}
\end{figure*}

%

\section{Case study: Cooperative Caching and Cache-Aided NOMA-MEC in NOMA-F-RANs}\label{section:casestudy}
In this section, two case studies of AI-driven feature extraction and cooperative caching in NOMA-F-RANs are presented to show the validity of our devised AI solutions.

\subsection{AI-driven Latent Feature Extraction in NOMA-F-RANs}

In this subsection, we first evaluate the performance of our proposed LSTM solution for F-UEs' task popularity prediction. More specifically, we formulate task popularity as a time series and treat the history task popularity and future popularity as the input time series and output time series of LSTM network, respectively.

Figure~\ref{performanceofLSTM2} demonstrates the performance of LSTM solution for task popularity prediction. The training dataset of task popularity are generated in a random walk model. The length of the generated task popularity is 500, which is corresponding to 500 time slots. The value of the task popularity in each time slot is between 0 and 1. The input length of the LSTM is set as 5, therefore, we have 455 pairs of data that can be used for training the LSTM networks. Moreover, we randomize the order to improve the generalization ability of the network. The inputs are task popularity data within 5 time slots and the labels are the task popularity in 6th time slot. $70\% $ of the task popularity data are utilized for the training process and $30\% $ of the task popularity data are applied for the testing process. In Fig.~\ref{performanceofLSTM2}, the ``real data" is the task popularity we generated with the random model, and the ``predicted data" is the predicted task popularity from LSTM networks. The goal in Fig.~\ref{performanceofLSTM2} is defined by mean square error (mse) between the target output and the network prediction result. The value of goal is selected according to experience. Higher goal brings more accuracy for content popularity prediction, yet may results in overfitting. According to Fig.~\ref{performanceofLSTM2}, we observe that the prediction error is marginal, which demonstrates that the proposed LSTM-based prediction framework can well estimate tasks' popularity. Moreover, the proposed LSTM solution outperforms conventional RNNs. In addition, a better performance can be obtained when we reduce the value of the goal.

\subsection{AI-driven Cooperative Caching in NOMA-F-RANs}

In the considered cooperative caching networks, there is a circle cooperative area composed of multiple BSs, F-UEs and massive wireless contents. Each BS contains limited caching capacity. The cooperative caching policy is to determine ``which" contents should be cached in ``which" BS. During the content delivery process, NOMA technique is utilized for sending the contents to the F-UEs simultaneously, thus improve the quality-of-experience (QoE) of F-UEs. In the considered simulation environment, the length of the cooperative region is 4km. There are multiple F-APs and F-UEs in the considered cooperative region. The F-UEs are randomly distributed in this cooperative region. For the whole network, there are 10 contents that the users may request. The workstation for the simulation has an Intel Core i7-7700 3.6 GHz CPU and 16 GB memory. The details of the network setting can be found in~\cite{zhong2019arxivcaching}.

In this subsection, a LAQL algorithm is proposed for cooperative caching, where the state space and action space are discrete. The details of the LAQL solution for cooperative caching in NOMA-F-RANs are given in~\cite{zhong2019arxivcaching}. The schematic diagram of LA-based Q-learning for NOMA-F-RANs is given in Fig.~\ref{Qlearningflow} that consists of three steps. In Step 1, an action probability vector is maintained. The action of the intelligent is chosen utilizing the probability distribution of the action space. Then, a feedback is calculated after taking this action. Then, according to the calculated reward/penalty, we estimate the maximum likelihood reward probability and choose the optimal action for the current training process. After the training process, the action probability vector can converge to a stable optimal vector, and then the optimal action reveals itself simultaneously in Step 3.

\begin{figure} [htp]
\setlength{\abovecaptionskip}{-0.2cm}
\centering
\includegraphics[width=3in]{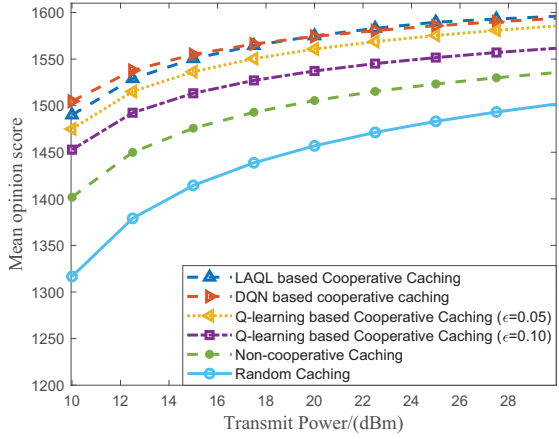}
 \caption{The MOS of different cache schemes vs. transmit power in AI-driven NOMA-F-RANs.}\label{Qlearningcaching}
\end{figure}

Fig.~\ref{Qlearningcaching} shows the mean opinion score (MOS), which is utilized for measuring the QoE of F-UEs in NOMA-F-RANs. Firstly, to avoid the significant influence of the learning rate $\alpha$ and the discount factor $\gamma$ on the convergence performance of the proposed LAQL, we regulate $\alpha$ and $\gamma$ for Q-learning. Then, after several adjustments, we set $\alpha$ as 0.75 and $\gamma$ as 0.6. Thereinafter, the learning rate and the discount factor of LAQL are set to be the same as that in conventional Q-learning, which is capable of eliminating the learning rate and the discount factor affect. It can be seen that LAQL based cooperative caching outperforms considered benchmark schemes, including conventional Q-learning, non-cooperative caching scheme and random caching scheme. The reason for this observation is that the designed action selection scheme of LAQL enables the intelligent agent to filter the optimal action for each given state, while the action selection strategies for conventional RL algorithms are designed by stochastic mechanisms, such as $\epsilon$-greedy. Fig.~\ref{Qlearningcaching} also demonstrate that the proposed cooperative caching framework outperforms conventional Q-learning based solution and random caching by 0.94\% and 4.72\%, respectively, corroborating that the designed cooperative caching scheme is capable of better balancing the caching capacities for all F-APs compared to conventional non-cooperative caching.
\vspace{-0.3cm}

\section{Conclusion Remarks and Future Challenges}\label{section:Conclusion}

In this article, the application of NOMA technique in F-RANs has been exploited. The network structure of NOMA-F-RANs is demonstrated where the NOMA technique is adopted to accommodate multiple F-UEs in a single resource block. Then, key modules in NOMA-F-RANs, named latent feature prediction, cooperative caching and cache-aided MEC have been studied in detail. The AI solutions for resources allocation problems in NOMA-FRANs have been identified, followed by two case studies, using LSTM for task popularity prediction and LAQL for content placement in cooperative caching, respectively. There are still many open research problems in this area, which are outlined as follows:
\begin{enumerate}
  \item[$\bullet$] {\bf Privacy~and~security~preserving:} The advantages of AI algorithms in NOMA-F-RANs rely on the data that can be subject to F-UEs' privacy and security. The data of the F-UEs is maintained for the intelligent agent to implement the AI algorithms. If the data is leaked by a data breacher, then both the privacy and security of F-UEs are compromised.

  \item[$\bullet$] {\bf Network~edge~caching~reaping:} The thriving of mobile F-UEs create a large amount of caching capacity. Reaping the benefits of caching resources across many F-UEs as a whole has significant impact on the performance of NOMA-F-RANs. Some initial research contributions on D2D communication have been conducted to transmit the data between different F-UEs, which leads to the more practical but challenging data cooperative caching. Such cooperative caching designs still constitute an open area.

  \item[$\bullet$] {\bf Other~proactive~schemes~to~boost~the~gain:} The caching efficiency of the AI enabled cooperative caching scheme is reduced when the F-UEs' content request probabilities are highly heterogeneous. To overcome this shortage, some other proactive caching schemes have been proposed to improve the diversity of content popularity. By recommending some popular contents, the recommendation system can reshape F-UEs' content request distributions, resulting in boosted cache hit ratio. Proactive schemes should be jointly designed to further enhance the performance gains of NOMA-F-RANs.

  \item[$\bullet$] {\bf Other~key~challenges~when~applying~AI~to~NOMA-FRANs:} There are other critical challenges when applying AI to NOMA-F-RANs, such as the change in network size and long convergence time, which still need great effort. Fortunately, some progresses have been made, such as meta-transfer learning (MTL) approaches which are capable of adapting DNNs for few shot learning models. The key point of MTL is to train multiple tasks, and learn scaling and shifting functions of network weights for each model. In addition, federated learning approaches are proposed for improving the convergence of DNNs, by training multiple local models and updating global model using well-trained local models.
\end{enumerate}

\linespread{0.85}
\bibliographystyle{IEEEtran}
\bibliography{mybib}

 \end{document}